\documentclass[12pt]{article}
\usepackage{graphicx}
\usepackage{cancel}[]
\usepackage{amssymb,amsmath}
\usepackage{ifpdf}
\usepackage{hyperref}[]
\usepackage{fullpage}
\usepackage{color}
\usepackage{ulem}
\usepackage{verbatim}
\usepackage{float}
\ifpdf
    \DeclareGraphicsExtensions{.pdf}
\else
    \DeclareGraphicsExtensions{.eps}
\fi

\usepackage{graphicx,epsfig}
\usepackage{cancel}[]
\usepackage{amssymb,amsmath}
\usepackage{ifpdf}
\usepackage{hyperref}[]
\def\lsim{<\kern-2.5ex\lower0.85ex\hbox{$\approx$}\ }
\def\rsim{>\kern-2.5ex\lower0.85ex\hbox{$\approx$}\ }

\def\LAMBDABAR
{\hbox{$\lambda$\kern-0.52em\raise+0.45ex\hbox{--}\kern+0.2em}}

    \DeclareGraphicsExtensions{.eps}

\begin{document}

 \centerline{\Large\bf{Upper limit on the Stiffness of space-time }}
%\centerline{\Large\bf{as deduced from the propagation of gravitational waves }} 
\vspace{.20in}
%\end{document}
%\vspa

 \centerline{A. C. Melissinos }
\centerline{\it{Department of Physics and Astronomy, University of Rochester }}
\centerline{\it{Rochester, NY 14627-0171, USA}}
\vspace{.20 in}
%\centerline{  May 30, 2018}
%\end{document}
\vspace{.30 in}
\begin{footnotesize}
{\it{Abstract}} From the recently observed propagation of gravitational waves through space-time \cite{GW} 
an upper limit can be deduced for the stiffness of space-time through which the gravitational waves propagate.
This upper limit is extremely weak, implying that the stiffness of space-time is at least 14 orders of magnitude 
weaker than that of jello.
\end{footnotesize}
\vspace{.20 in} 
%\end{document}

At the recent 2018 Hamilton Lecture at Princeton ``Exploring the Universe with
Gravitational Waves" the question arose as to what is the Young's modulus $Y$ of 
space-time \cite{Kirk} as experienced by a gravitational wave (GW). One response 
was that $Y_{space-time}$ is frequency
dependent and given by \begin{equation} Y_{space-time} \approx c^2f^2/G, 
\end{equation} where $f$ is the frequency of the gravitational wave, and $G$
the Newtonian Gravitational constant. This 
relation is reached by equating the energy density in the gravitational wave,
see for instance Eq.(107.11) in Landau-Lifschitz \cite{Landau} or Eq.(10.3.6) in
Weinberg \cite{Weinberg},
\begin{equation} u_{GW} = \frac{\pi c^2 f^2 h^2}{8G} \end{equation} 
to the mechanical energy density of a loaded bar with 
Young's modulus $Y$, and subject to the same strain  $h$ as imposed by the GW
\begin{equation} u_{mechanical} = Yh^2/2,\end{equation} see for instance
\cite{MandL} or any elementary text on the strength of
materials.  This leads to
\begin{equation} Y_{space-time} = \frac{\pi c^2 f^2}{4G}\end{equation}
as quoted in Eq.(1) above.\\

We disagree with this interpretation because it uses a mechanical analogy of
space-time, and prefer instead to work from first principles. 
Consider a GW in the TT gauge, propagating along the z-axis and a small segment
of space normal to the direction of propagation with thickness $dz$ and sides
of length $L$ along the $x$ and $y$ directions. 

\newpage
For a wave polarized in the $h_{+}$ mode the acceleration imposed on a differential 
volume of the medium are, see for instance Eq.(1.98) and ff. in \cite{Maggiore}, 
or Eq.(37.6) of \cite{MTW}
\begin{equation}\delta \ddot{x} = -\frac{h_{+}}{2}x \omega^2 {\rm{sin}}\omega t
\end{equation} \begin{equation}
\delta \ddot{y} =\ \frac{h_{+}}{2}y \omega^2 {\rm{sin}}\omega t \end{equation}
At $\omega t = \pi/2$, the tidal accelerations of Eqs.(5,6) exert a compressional 
force in the x-direction
and an elongation in the y-direction. The mass of the thin slab is \begin{equation}
M= \rho_{mass}L^2 dz = (\rho/c^2) L^2 dz\end{equation}
where $\rho$ is the energy density of space-time. 
Thus the compression force (along the x-direction) is\begin{equation}
F_x = M \delta \ddot{x} = M (2\pi f)^2 h_{+} L = (\rho/ c^2) (2\pi f)^2 h_{+} L^3 dz.
\end{equation} Similar results obtain for the y-direction.
 
The stress in the $x$ direction, on the space-time volume, is $ S_x = F_x/A = F_x/L dz $ 
and by definition the Young's modulus
\begin{equation} Y = \frac{S_x}{h_{+}} = (\rho/ c^2) (2\pi f L)^2 \end{equation}
The extent, $L$, of the test volume depends on the curvature of the wavefront of the 
gravitational wave, but there is an upper limit that can be set. For the quasi-Newtonian
discussion adopted here, the force exerted by the gravitational wave in Eq.(8), is valid
  provided $L\ll\lambda_{GW}/2\pi = c/(2\pi f)$ \cite{Maggiore}. Therefore 
\begin {equation} Y \ll \rho \end{equation}
We thus find that the Young's modulus of space-time is frequency independent 
and is limited by
the energy density of the matter through which the GW propagates. For cosmic 
distances, such as traversed by GWs it is appropriate to use the critical
density of the universe
%\end{document}
$$ \rho_{c}/c^2 = 10^{-29}\ {\rm {g/cm^3}} \qquad {\rm{hence}} \qquad \rho_{c}\approx 10^{-9} 
{\rm {J/m^3}} = 10^{-9}\ \rm{Pa}.$$
Thus we conclude that that the stiffness of space-time is 14 orders of magnitude
{\it{less}} than that of jello (for which $Y\approx 10^5$\ Pa).\\
%\end{document}

It is also interesting to estimate the energy loss of the GW as a
function of the Young's modulus, as it propagates from its origin to our
detectors on Earth. In traversing a material medium the GW displaces   from
their equilibrium position differential elements of the material
(transversly to its direction of propagation),
see Eqs.(5,6). The displacements $u_x$ and $u_y$ are
%\end{document}

\begin{equation} u_{x}(t) =\ \frac{h_{+}}{2} x_{0} {\rm{sin}}\omega t \qquad 
{\rm{and}} \qquad 
u_{y}(t) = -\frac{h_{+}}{2}y_{0} {\rm{sin}}\omega t \end{equation}
%\end{document}
The energy for such oscillations of the material is provided by the GW which
is subject to energy loss 
\begin{equation} E =\int \frac{1}{2}\left[ \left(\frac{\partial u}
{\partial t}\right)^2 + v_s^2  \left(\frac{\partial u}
{\partial x}\right)^2 \right] dm \end{equation} where \begin{equation}
v_s = c \sqrt{Y/{\rho}} \end{equation} is the velocity of sound in the medium.
The energy dissipated by the GW when traversing through a volume
$V= A \Delta z$ is
\begin{equation} E_d = \frac{\rho}{c^2}\frac{A}{2} \int[h_{+}^2 \omega^2 x_0^2 {\rm{sin}}^2 \omega t 
+ v_s^2 h_{+}^2 {\rm{cos}}^2 \omega t] dz =  \frac{\rho A \Delta z}{2}\left(\frac{v_s}{c}\right)^2 h_{+}^2=
\frac{A \Delta z}{2} h_{+}^2 Y \end{equation}
We used the fact that the two terms in Eq.(12) are equal representing the kinetic
and potential energy of the material oscillations, and averaged over the time
dependence. The total path length traversed by the GW is designated by $\Delta z$.\\

The energy stored in the GW is obtained by multiplying the energy density in the
wave, Eq.(2) by the volume $V = A( c \Delta \tau)$, where $\Delta \tau $ is the length
of the GW burst, \begin{equation} E_{GW} = A ( c \Delta \tau) \frac{\pi c^2 h_{+}^2 f^2}{8G}\end{equation}
Clearly $E_{GW}$ must exceed the dissipated energy $E_d$ leading to
\begin{equation} Y < \frac{\pi c^2 f^2}{4 G} \frac{c \Delta \tau}{\Delta z}\end{equation}
Using from ref. \cite{GW} $\Delta \tau \approx 1\ {\rm{s}}$ and $\Delta z \approx 400$\ Mpc,
%and setting $f= 100$ Hz, as in \cite{Kirk},
we find that  
\begin{equation}  Y < 2.5\times 10^{-17}(c^2 f^2/G) \end{equation}
This contradicts Eq.(1 or 4), but is fully compatible with Eq. (10).\\

For some recent references to the extensive literature on the stiffness of space-time see
for instance \cite{ Tenev, Izabel}.

\vspace{0.2 in}
%\section{Acknowledgment}
{\bf\Large{Acknowledgement}}

I thank Professor Kirk T. McDonald for bringing this issue to my attention and for
sharing his unpublished note \cite{Kirk}, as well as for extensive discussions.

\end{document}